\documentclass[aps,pra,
twocolumn,
unsortedaddress,superscriptaddress,
]{revtex4-1}

\usepackage{graphicx}% Include figure files
\usepackage{dcolumn}% Align table columns on decimal point
\usepackage{bm}% bold math
\usepackage{tabularx}
\usepackage{physics}
\usepackage{bm}%
\usepackage{color}      % use if color is used in text
\usepackage[colorlinks=true,linkcolor=blue,citecolor=blue,urlcolor=blue]{hyperref}%
\expandafter\ifx\csname package@font\endcsname\relax\else
 \expandafter\expandafter
 \expandafter\usepackage
 \expandafter\expandafter
 \expandafter{\csname package@font\endcsname}%
\fi
\begin{document}

\author{Bin-Bin Xu}
\affiliation{%
Centre of Excellence for Quantum Computation and Communication Technology, School of Physics, University of New South Wales, NSW 2052, Australia
}

\author{Gabriele G. de Boo}
\affiliation{%
Centre of Excellence for Quantum Computation and Communication Technology, School of Physics, University of New South Wales, NSW 2052, Australia
}

\author{Brett C. Johnson} % https://orcid.org/0000-0002-2174-4178
\affiliation{%
Centre of Excellence for Quantum Computing and Communication Technology, School of Physics, University of Melbourne, Victoria 3010, Australia
}%

\author{Milo\v{s} Ran\v{c}i\'{c}} % https://orcid.org/0000-0002-8672-6771
\affiliation{%
Centre of Excellence for Quantum Computation and Communication Technology, Research School of Physics and Engineering, Australian National University, ACT 0200, Australia
}%

\author{Alvaro Casas Bedoya}% https://orcid.org/0000-0002-0947-630X
\affiliation{%
Institute of Photonics and Optical Science (IPOS), School of Physics, The University of Sydney, NSW 2006, Australia
}%
\affiliation{%
The University of Sydney Nano Institute (Sydney Nano), The University of Sydney, NSW 2006, Australia
}%

\author{Blair Morrison}% https://orcid.org/0000-0002-7150-1887
\affiliation{%
Institute of Photonics and Optical Science (IPOS), School of Physics, The University of Sydney, NSW 2006, Australia
}%
\affiliation{%
The University of Sydney Nano Institute (Sydney Nano), The University of Sydney, NSW 2006, Australia
}%
\author{Jeffrey C. McCallum} % https://orcid.org/0000-0002-6692-7728
\affiliation{%
Centre of Excellence for Quantum Computing and Communication Technology, School of Physics, University of Melbourne, Victoria 3010, Australia
}%

\author{Benjamin J. Eggleton}% https://orcid.org/0000-0003-4921-9727
\affiliation{%
Institute of Photonics and Optical Science (IPOS), School of Physics, The University of Sydney, NSW 2006, Australia
}%
\affiliation{%
The University of Sydney Nano Institute (Sydney Nano), The University of Sydney, NSW 2006, Australia
}%

\author{Matthew J. Sellars} % https://orcid.org/0000-0002-0502-0444
\affiliation{%
Centre of Excellence for Quantum Computation and Communication Technology, Research School of Physics and Engineering, Australian National University, ACT 0200, Australia
}%

\author{Chunming Yin}% https://orcid.org/0000-0003-0117-8225
 \email{c.yin@unsw.edu.au}
\affiliation{%
Centre of Excellence for Quantum Computation and Communication Technology, School of Physics, University of New South Wales, NSW 2052, Australia
}

\affiliation{%
CAS Key Laboratory of Microscale Magnetic Resonance and Department of Modern Physics, University of Science and Technology of China, Hefei 230026, China
}
\author{Sven Rogge} % https://orcid.org/0000-0003-1078-9482
\affiliation{%
Centre of Excellence for Quantum Computation and Communication Technology, School of Physics, University of New South Wales, NSW 2052, Australia
}

\title{Ultra-shallow junction electrodes in low-loss silicon micro-ring resonators}

\begin{abstract}
Electrodes in close proximity to an active area of a device are required for sufficient electrical control. The integration of such electrodes into optical devices can be challenging since low optical losses must be retained to realise high quality operation. Here, we demonstrate that it is possible to place a metallic shallow phosphorus doped layer in a silicon micro-ring cavity that can function at cryogenic temperatures. We verify that the shallow doping layer affects the local refractive index while inducing minimal losses with quality factors up to 10$^5$. This demonstration opens up a pathway to the integration of an electronic device, such as a single-electron transistor, into an optical circuit on the same material platform.
\end{abstract}

\maketitle

%%%%%%%% 
%%%%%%%% 
\section{Introduction}
	Electrical readout of impurity spin states can be achieved on a Si platform and is the basis of qubit devices for quantum computer and communication applications \cite{Veldhorst:2015aa,He:2019aa, Borjans:2020aa,Samkharadze1123}. The success of this system is owed to the high purity of the material, which results in the remarkably long quantum coherence times \cite{Muhonen2014, Veldhorst:2014aa}, and to the extremely well-developed Complementary Metal-Oxide-Semiconductor (CMOS) technology. Silicon on insulator (SOI) is also an attractive material for transmission of near infrared (NIR) photons due to the low scattering and absorption loss. These photons may be encoded with quantum state information of an incorporated impurity and, in this way, is a promising route for the long range coupling between qubits. To achieve this, both optical and electrical access to the qubit is required.

	Active ring resonators consisting of doped resonator cores have been investigated for use as electro-optic modulators \cite{Dong:09}. Modulation of the transmitted signal is achieved through changing the carrier density within the ring which in turn modulates the refractive index \cite{Xu:2005aa}. The fastest modulation times can be achieved if a $pn$ junction is formed within the core of the ring, although this can have a negative impact on the Q factor, defined as $\lambda / \Delta \lambda$ where $\lambda$ is the wavelength \cite{Dong:09}. Dopant implantation has been used to introduce junctions into the resonator with great success. This approach is compatible with Si-based qubit device formation using the top-down approach.

	Here, we design a CMOS compatible process to integrate electrodes directly on a ring cavity fabricated on a SOI platform for eventual control and read-out of qubit spin states. The electrodes are formed via ion implantation of phosphorus (P). The doped area in the semiconductor is only a few Bohr orbits thick, i.e. less than 20 nm. This would allow a qubit to be addressed both optically and electrically simultaneously. The electric field created by the electrodes can be localised to a small region within the ring cavity. The effect of such electrodes on the optical cavity modes is explored. Large quality factors are shown to be retained after processing. Simulations of the cavity - electrode system is simulated and show good agreement with experimental results. Electrical transport measurements under conditions required for qubit control are also demonstrated. Such structures may be promising platforms for the control of qubits in other materials such as the divacancy or silicon vacancy in SiC \cite{Castelletto:2020aa}. This work represents the first step towards realising strong optical coupling compatible with silicon-based quantum computing architectures.

%%%%%%%% 
%%%%%%%% 
\section{Methods}

%%%%%%%%%%%%%%%%%%%%%
%%%%%%%%%%%%%%%%%%%%%
%%%%%%%%%%%%%%%%%%%%%
\begin{figure}
\begin{center}
 \rotatebox{0}{\includegraphics[width=8.5cm]{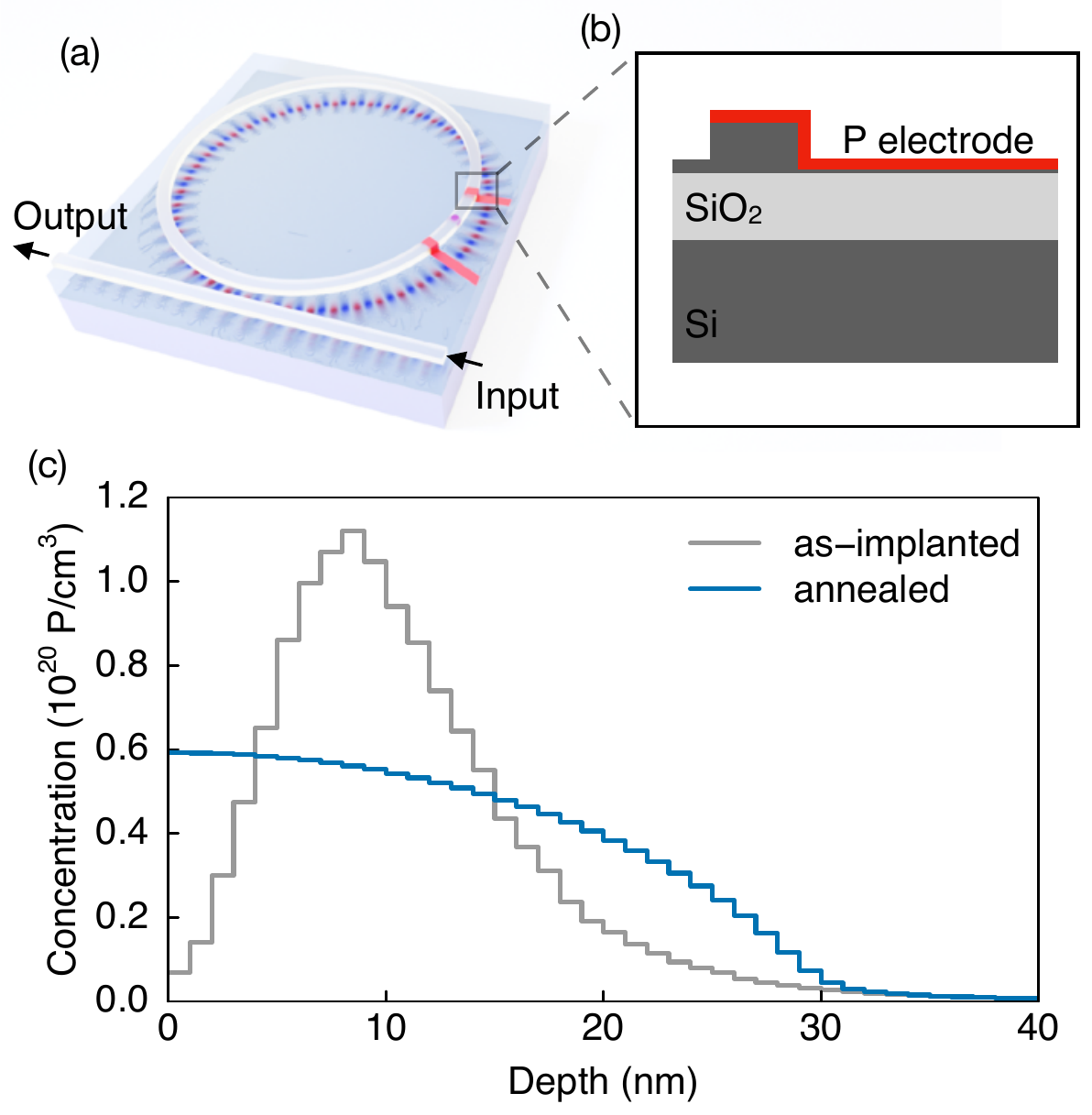}}
\end{center}
\caption[]{
(a) Conceptual schematic of the device, which consists of a Si ring resonator and integrated cryogenic-compatible electrodes for qubit read-out. (b) A cross-section of the ring showing the profile of the electrode. (c) A Monte Carlo simulation of the implanted 4~keV P profile. In this example the implantation fluence was $1.3\times 10^{14} \;\rm cm^{-2}$ resulting in a peak P concentration of $1.1\times 10^{20} \;\rm cm^{-3}$. The expected P concentration after annealing for 15~s at 900$^{\circ}$C is also shown.
} 
\label{F1}
\end{figure}

Figure~\ref{F1}(a) shows a conceptual image of a micro-ring cavity and evanescent field coupled bus waveguide. Integrated P doped electrodes for electrical control of qubits situated within the ring are shown in red. For this study, the whispering gallery mode (WGM) micro-ring cavities were fabricated at IMEC and have diameters of 20 and 34.3~$\mu$m. The cavities are formed on a silicon-on-insulator (SOI) wafer consisting of a 220~nm thick top Si layer, a 2~$\mu$m thick SiO$_2$ layer and an 800~$\mu$m Si substrate. The width of the micro-rings is 450~nm.

Photons are coupled into waveguides with a single mode fibre via grating couplers at an angle of 10$^\circ$ to the normal direction. The gratings are designed for a central wavelength of 1550~nm, ideal for impurities like Er \cite{Yin:2013aa}, with a period of 630~nm and depth of 70~nm. These are also polarization sensitive and optimized for TE polarized light. The electric field profile of light propagating from the input to the output ports via a waveguide is also shown in Fig.~\ref{F1}(a). 

Optical transmission through the ring cavities is measured with a fibre-coupled tunable laser source with a wavelength resolution of 1~pm. Spectra were acquired in the 1510 to 1570~nm range which is compatible with Er$^{3+}$ absorption \cite{Yin:2013aa}. The input and output light intensity was monitored to determine the transmission properties and resonant frequencies of the ring cavity. The total loss includes the coupling and transmission loss.

To form the electrodes, phosphorus (P) implantation was employed. This is an industry compatible technique for shallow junction formation. The implantation is performed directly into the micro-ring cavity in close proximity to where an individual impurity/defect may be addressed both optically and electrically. The region surrounding the ring consists of a thin layer of Si in which the electrode is also patterned. A cross-section of the P doped lead is shown in Fig.~\ref{F1}(b). Aluminium contact pads are connected to these leads. 

A number of different implant strategies were considered. The micro-ring cavity was either entirely implanted, locally implanted over a 200~nm length or implanted in several locations to form actual electrodes. The P implant window was defined in a PMMA surface layer with a Raith 150 electron beam lithography (EBL) system. The implant was achieved with a P$_2$ molecular beam accelerated to 8~keV (4~keV per P ion). These implantation parameters result in a shallow P profile with a projected range of 8.5~nm according to Monte Carlo simulations performed with crystal-TRIM (Fig.~\ref{F1}(c)) \cite{crystalTRIM}.  The P concentration is in the degenerate doping regime to ensure they function at cryogenic temperatures.
 
After implantation, the PMMA was removed with an O$_2$ etcher. The P was then electrically activated with an optimized rapid temperature anneal (RTA) at 900$^\circ$C for 15~s in an Ar ambient. The expected P concentration profile after diffusion at this temperature is also shown in Fig.~\ref{F1}(c), showing the high P concentration at the Si surface. The device with the electrodes received a further EBL step to define Al contact pads deposited by physical vapor deposition. To enhance the adherence and Ohmic quality between the implanted P and Al, the device was annealed for 10~mins at 400$^\circ$C. This device was then fixed to a PCB board designed for conductivity testing. A four probe source meter was used to measure the current-voltage characteristics of the doped area within the micro-ring resonator.

%%%%%%%% 
%%%%%%%% 
\section{Results and discussion}

%%%%%%%%%%%%%%%%%%%%%
%%%%%%%%%%%%%%%%%%%%%
%%%%%%%%%%%%%%%%%%%%%
\begin{figure}
\begin{center}
\rotatebox{0}{\includegraphics[width=8.5cm]{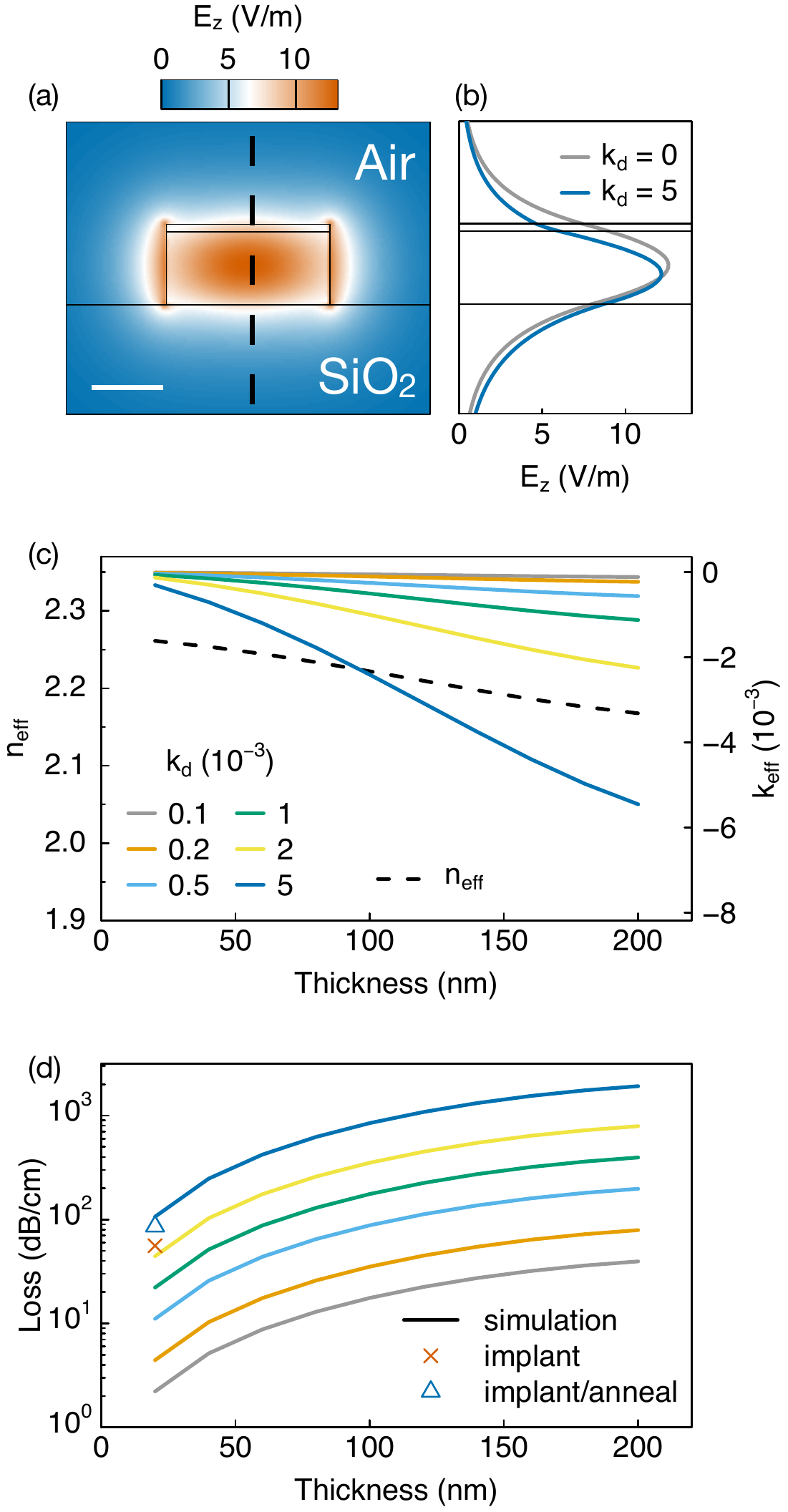}}
\end{center}
\caption[]{(a) Simulation of the electric field for a TE mode at a wavelength of 1550~nm inside a waveguide with a $220 \times 450\;\rm nm^2$ cross-sectional area. The scale bar is 0.2~$\mu$m (b) The electric field profile though the waveguide determined at the dashed line in (a). This is shown for top layers with $k_\mathrm{d} = 0$ and $k_\mathrm{d} = 5$. (c) The effective refractive indices $n_\mathrm{eff}$ and i$k_\mathrm{eff}$ for different $k_\mathrm{d}$ values and different doped layer thicknesses. The dashed line and the coloured lines are plotted against the left and right axes, respectively. (d) The corresponding loss calculated from the $k_\mathrm{eff}$ values in (c). Two experimental values for the loss (symbols) are also included as discussed in the text.
} 
\label{F2}
\end{figure}

To study the influence of the doped area on the optical cavity transmission further, we consider a simplified 2D model of a Si waveguide. The electric field distribution is shown in Fig.~\ref{F2}(a). The waveguide is 220~nm thick and 450~nm wide, with a SiO$_2$ substrate and air as the top cladding. The doping in the top 20~nm of the Si waveguide is assumed to be uniform. This is incorporated into the model via changes in the complex refractive index, $(n+ik)_\mathrm{d}$, with dopant concentration. Both parts of the refractive index depend on the doping concentration but the optical extinction coefficient, $k_\mathrm{d}$, related to the absorption efficient by the equation $\alpha = 2\pi k / \lambda$, is most strongly affected via free carrier absorption effects\cite{Soref:1987aa}. At a wavelength of $\lambda = 1550$~nm, $k_\mathrm{d}$ is approximately linear with P dopant concentration, $[P]$, $ k \propto 1.49\times 10^{-22} [P]$ \cite{Balkanski:1969aa}. 

The change in the resonant mode profile in the presence of the doped layer is shown in Fig.~\ref{F2}(b). This value, $k_\mathrm{d}=5$, is unrealistically high and is merely used to illustrate the effective shift of the mode deeper into the waveguide. In Fig.~\ref{F2}(c) and (d) the effective refractive index and loss dependence on $k_\mathrm{d}$ and the doped layer thickness is considered. Changes in the doped layer thickness can be achieved experimentally by implanting P at more than one energy to create a box-like concentration profile. Firstly, it is shown that both the real and imaginary parts of the effective refractive index for the principle mode decreases as the doped layer thickness increases. The loss is evaluated from $k_\mathrm{eff}$ and, as might be expected, is found to increase as $k_\mathrm{d}$ or the doped layer thickness increases. Two data points are included in Fig.~\ref{F2} as discussed further below. 

% For the case where a short section of the waveguide is implanted, no obvious mode mismatches around the doping area can be observed  for sufficiently low $k_\mathrm{d}$ values, as explored in 3D models included in the supplementary material S1.

%%%%%%%%%%%%%%%%%%%%%
%%%%%%%%%%%%%%%%%%%%%
%%%%%%%%%%%%%%%%%%%%%

\begin{figure}
\begin{center}
\rotatebox{0}{\includegraphics[width=8.5cm]{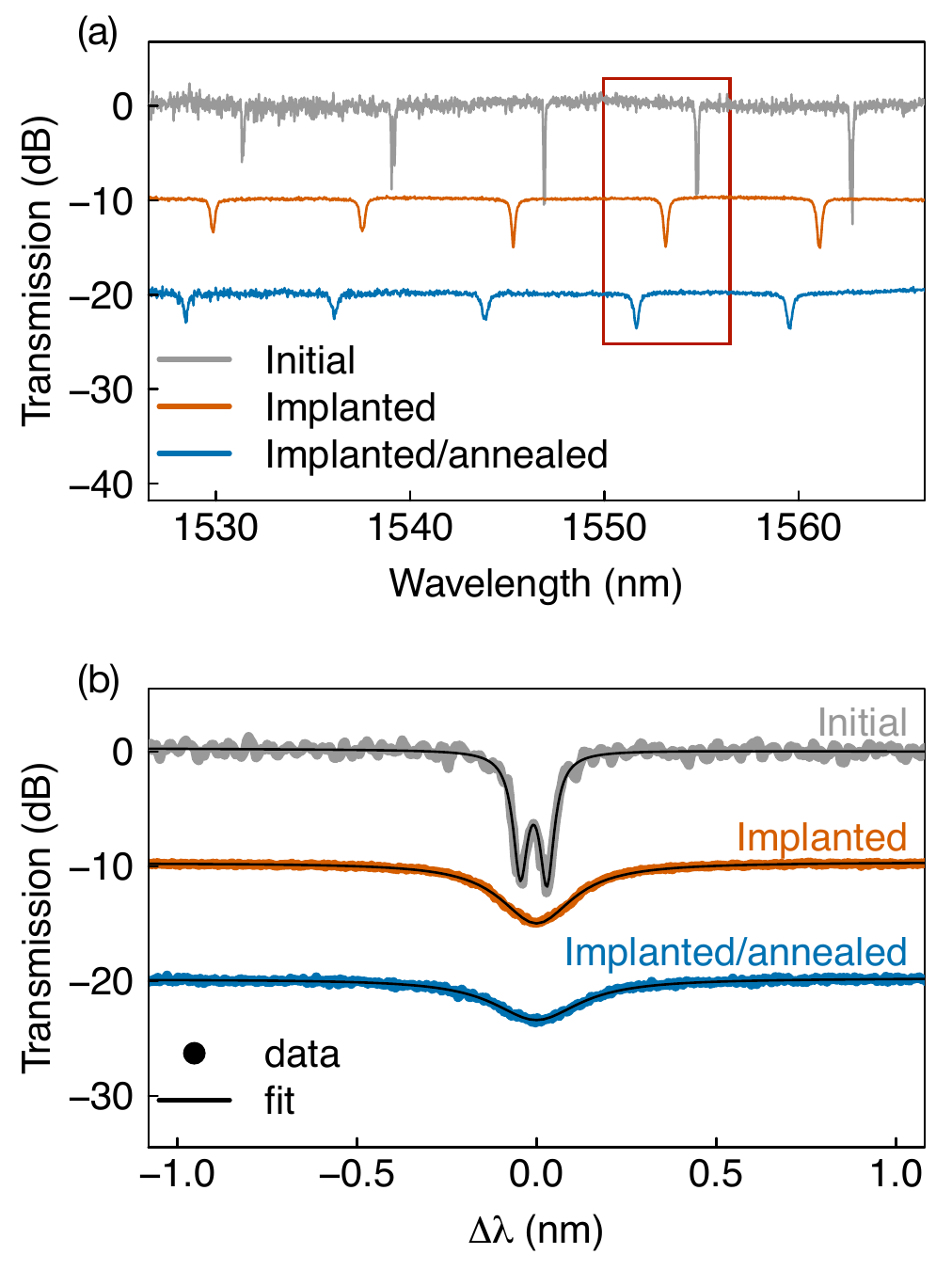}}
\end{center}
\caption[]{(a) Transmission spectra for a 20~$\mu$m diameter ring cavity measured before implantation, after implantation, and after post-implantation annealing. Spectra have been offset vertically by -10~dB for clarity. (b) Zoom-in of the region indicated by the box in (a). These spectra have been shifted by 1554.78, 1553.15, and 1551.65~nm, respectively. Solid lines are Lorentzian fits. The split resonance has been fitted with two independent Lorentzians.
} 
\label{F3}
\end{figure}

When the entire top surface of an actual ring cavity is P implanted and then annealed we see that the resonant mode wavelength and Q factor are modified as shown in the transmission spectra in Fig.~\ref{F3}. A zoom in of the modes around 1552~nm are shown in Fig.~\ref{F3}(b). Initially, the ring resonators exhibited excellent Q factors of $>2 \times 10^4$. A simple Lorentzian fit is used, which yields the total $Q$ that includes the coupling of the resonator to the waveguide \cite{An:2018} (complete fitting parameters for each resonant mode are included in the supplementary). With such high Q factors, resonant peak splitting of becomes apparent. In this case, the splitting is $\sim$0.07~nm as determined with two independent Lorentzians similar to \onlinecite{Li:2016}. These doublets are a consequence of contra-directional coupling where the degeneracy of forward and backward propagating modes within the ring resonator is lifted and they become coupled \cite{Little:97}. The extent of the splitting depends on the coupling strength and may be sensitive to surface roughness and indeed any post-fabrication processing steps. 

After implantation the resonant peak around 1554.78~nm, for example, shifts down by about 1.6~nm and the Q factor decreased from $3.2\times 10^4$ to $0.7\times 10^4$ due to absorption loss brought about by the implantation-induced damage and corresponding changes in the refractive index as described in Fig.~\ref{F2}. This then obscured the resonant mode splitting. After post-implantation annealing the resonant peak shifts further down by 1.5~nm and the Q factor dropped to $0.6\times 10^4$.

Based on the fitting results of all peaks in Fig.~\ref{F3}(b), the optical loss in the ring can be calculated \cite{An:2018} and compared between the three stages. The optical loss caused by implantation was estimated to be $56\pm6\;\rm dB/cm$, and the optical loss caused by implantation and annealing was estimated to be $86\pm10\;\rm dB/cm$. Initially, the low energy, high fluence implantation is expected to result in an amorphous silicon top layer with a corresponding increase in $k_\mathrm{d}$\cite{Luciani:1989}. During the post-implantation anneal the crystallisation occurs via solid phase epitaxy which can activate the implanted dopant well above the solid solubility limit \cite{Johnson2012}. This leads to stronger free carrier absorption effects. Concurrently, the anneal may also improve the surface roughness thereby enhancing the Q factors\cite{Zhu:09,Bourhill:19}. 
These two values are included in Fig.~\ref{F2}(d). According to the simulation, we see that the loss caused by the P implant/anneal process corresponds to $k_\mathrm{d} = 0.004$, or a dopant concentration of $2.7\times 10^{19}\;\rm P/cm^{3}$. This value is in fair agreement but lower than the expected P concentration of $6\times 10^{19}\;\rm P/cm^{3}$ (Fig.~\ref{F1}(c)). This under-estimation is most likely due to the assumption in the simulation that the top layer is uniformly doped. 

%%%%%%%%%%%%%%%%%%%%%
%%%%%%%%%%%%%%%%%%%%%
%%%%%%%%%%%%%%%%%%%%%

\begin{figure}
\begin{center}
\rotatebox{0}{\includegraphics[width=8.4cm]{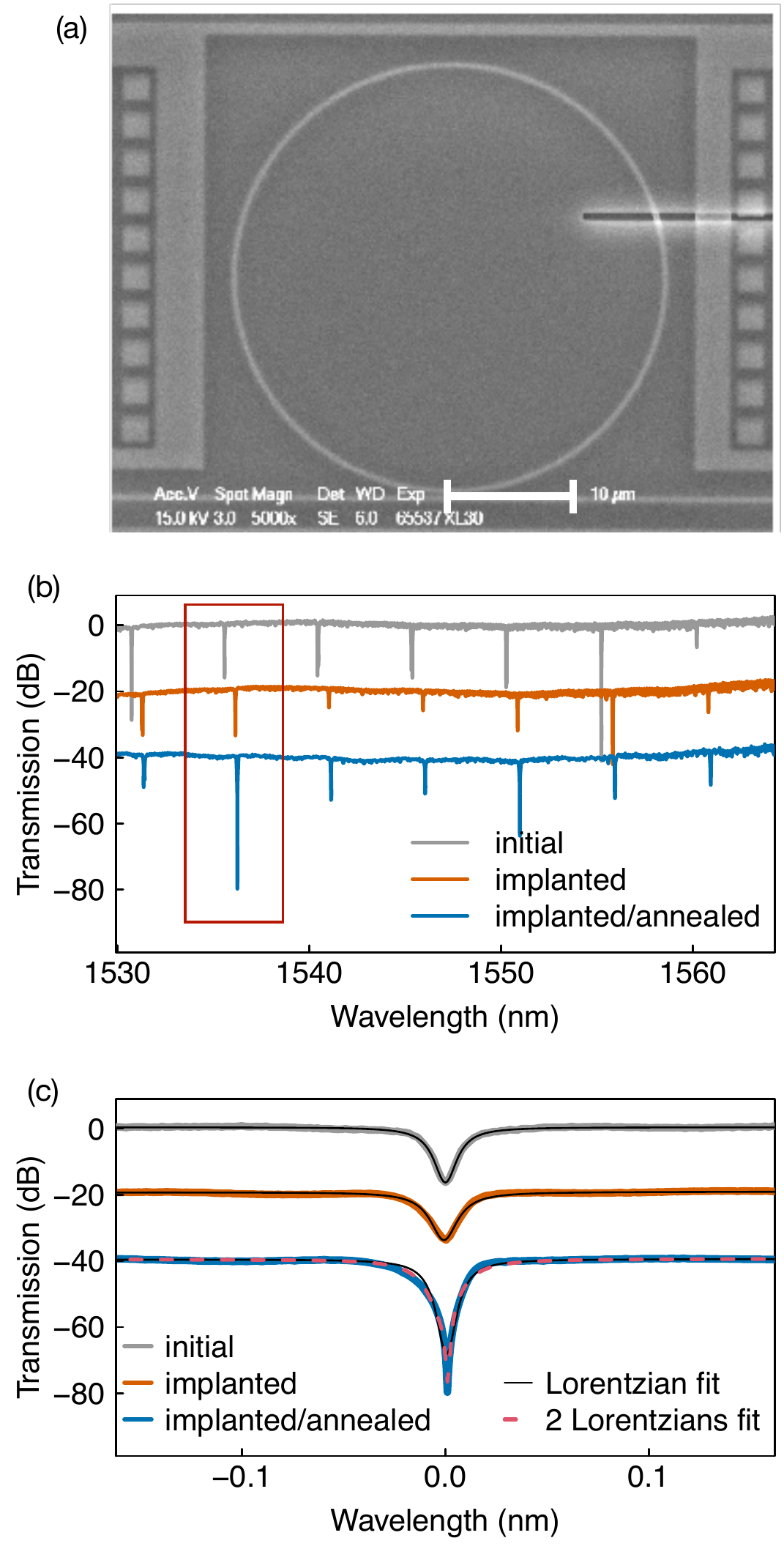}}
\end{center}
\caption[]{(a) An SEM micrograph of a 34.3~$\mu$m diameter micro-ring cavity doped in a $200 \times 450 \rm\; nm^2$ area (top-right region). The scale bar is 10~$\mu$m. (b) Spectra of Device 1 initially, after implantation and after the implant/anneal processing steps. The spectra are offset vertically by -20~dB for clarity. (c) A zoom-in of the resonant peaks around 1535.5~nm indicated by the box in (a). These spectra were shifted by 1535.59, 1536.15, and 1536.26~nm, respectively and fit with Lorentzians. Although resolved mode-splitting cannot be observed, a function consisting of two Lorentzians provides a better fit to the implanted/annealed device resonance in this case.   
}
\label{F4}
\end{figure}

We now discuss local P doping. Three ring-cavities, referred to as Device 1, 2, and 3, all with a diameter of 34.3~$\mu$m were selectively doped as shown in Fig.~\ref{F4}(a). The EBL defined window can be seen in the top right of the image. The transmission spectra of Device 1 taken at each stage of processing is shown in Fig.~\ref{F4}(b) over a wide spectral range. The FSR is about 5~nm and many of the modes again display splitting. A zoom in is shown for one set of modes around 1535.6~nm in Fig.~\ref{F4}(c). Contrary to the full implant case (Fig.~\ref{F3}) the resonant peaks shift upwards by about 0.41~nm after implantation and another 0.11~nm after dopant activation annealing. We also see an initial Q factor of $1.1 \times 10^5$ which decreases slightly to $0.96 \times 10^5$ on implantation. After the anneal, the Q factor increases to $1.2 \times 10^5$, greater than the initial value. However, for a number of the other modes the Q factor decreased. The full Lorentzian fitting parameters for all resonances are included in the supplementary material for completeness. This behaviour suggests that the process related to the implantation in general has a non-trivial impact on the behaviour of the ring resonator. In addition to changes in $k$ as a result of the activated dopant discussed above, the oxygen plasma used to remove the PMMA layer and the high temperature annealing process may modify the surface roughness. The latter may also modify the cross sectional shape of the waveguides slightly. In any case, we can conclude that it is possible to maintain high Q values with this implantation/anneal strategy.

%%%%%%%%%%%%%%%%%%%%%
%%%%%%%%%%%%%%%%%%%%%
%%%%%%%%%%%%%%%%%%%%%
\begin{figure}
\begin{center}
\rotatebox{0}{\includegraphics[width=8.5cm]{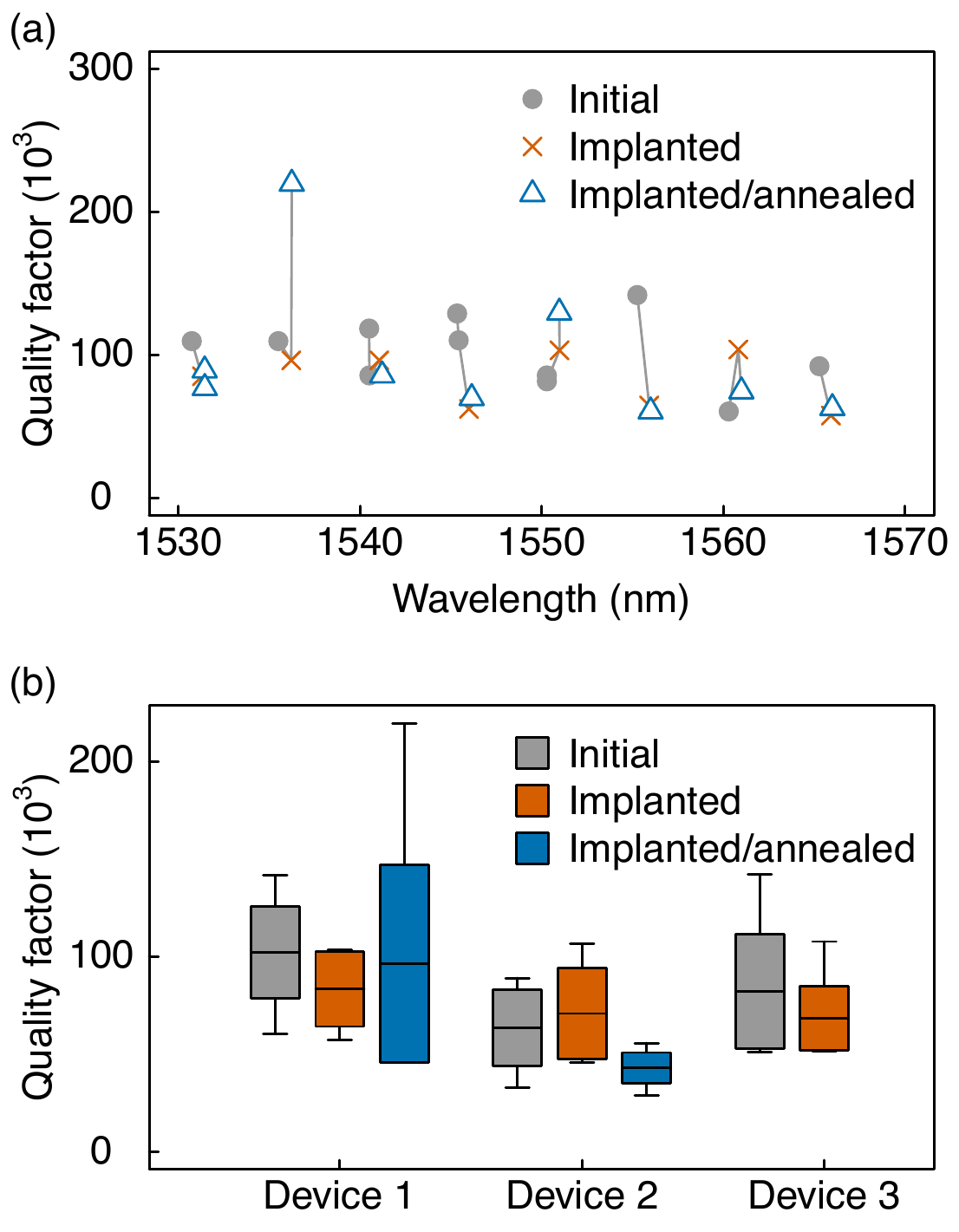}}
\end{center}
\caption[]{(a) The Q factor for Device 1 determined from the resonant peaks before implantation, after local (not full) implantation, and after annealing. (b) A boxplot of the Q factor distributions of all resonant peaks shown in (a) together with those of nominally similar devices within the 1530-1570~nm wavelength range. The vertical bounds of the boxes denote the standard deviation and the vertical lines the minimum and maximum values. The mean value of the Q factor is shown as a horizontal line across each box. The cavities of the three devices have the same diameter of 34.3~$\mu$m with an air gap of $150\pm 10$~nm.
} 
\label{F5}
\end{figure}

To evaluate the Q factor changes before and after local-area implantation and annealing, the Q factors for Device 1 are shown in Fig.~\ref{F5}(a). Like modes are connected via a solid line. A broad distribution of Q factors are apparent. This distribution is summarised for  the three devices in Fig.~\ref{F5}(b) where the variability in Q factor (and resonant wavelength not shown) most likely due to fabrication variations is clear. As observed for Device 1, the implantation results in a decrease in the mean Q factor value. The post-implantation anneal results in an increase of the mean close to the initial value. However, these changes are smaller than the statistical distribution of the Q values. This indicates that the integration of electrodes via shallow junction implantation is compatible with high Q values and that the required process steps do not increase the variability of the observed Q values compared to the undoped cavities. However, it does highlight the need for methods to tune the resonant wavelength that are compatible with operation at cryogenic temperatures.

%%%%%%%%%%%%%%%%%%%%%
\begin{figure}
\begin{center}
\rotatebox{0}{\includegraphics[width=8.5cm]{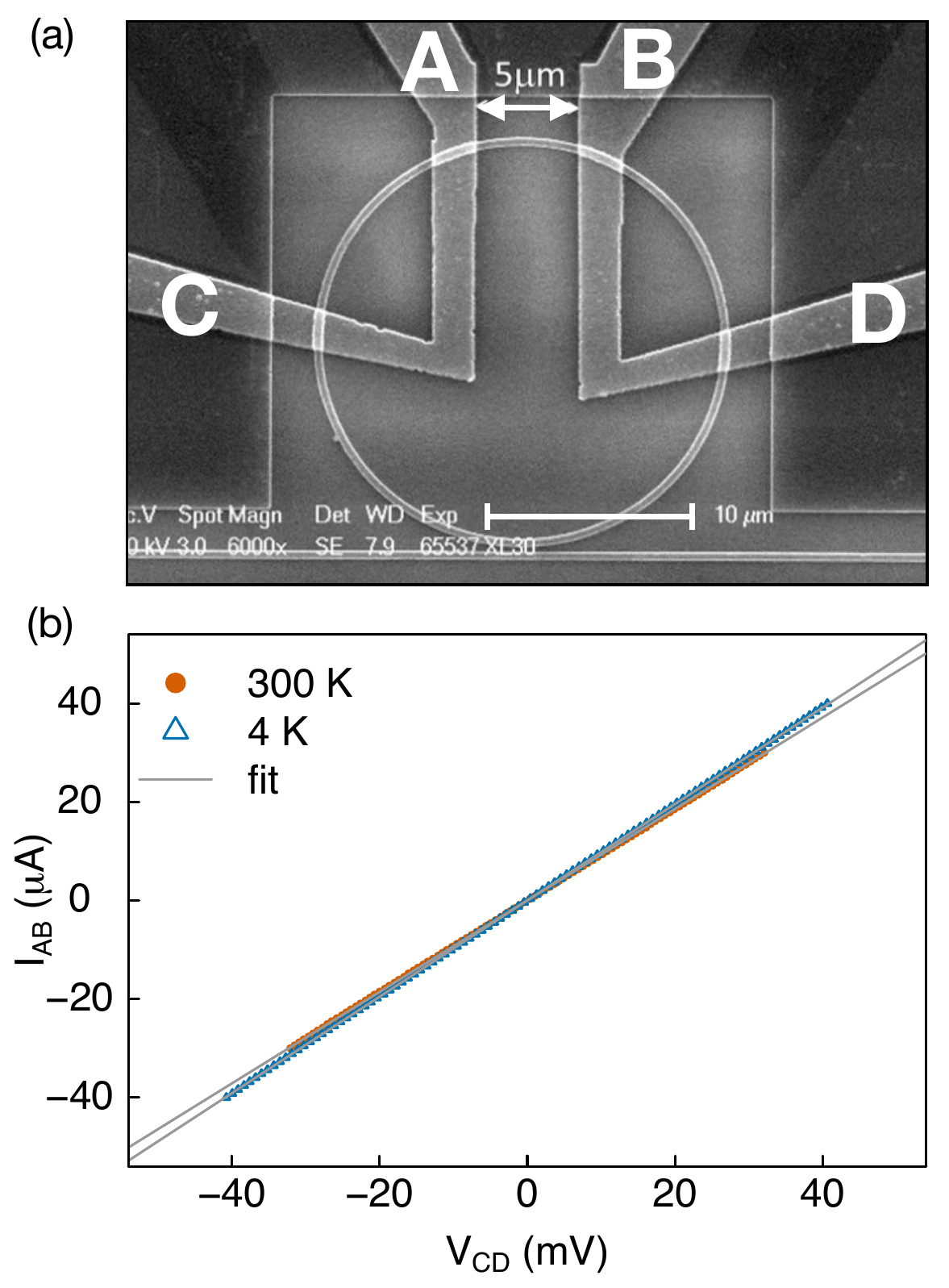}}
\end{center}
\caption[]{(a) SEM micrograph of a 14~$\mu$m diameter ring resonator device with a P doped section between contacts A and B. (b) The current-voltage traces measured at room temperature and at 4.2~K showing Ohmic behaviour as inferred by the linear fit.
} 
\label{F6}
\end{figure}

In this last section we demonstrate that the doped regions of the ring-cavity are indeed conducting at cryogenic temperatures and will therefore allow for qubit read-out. Fig.~\ref{F6}(a) shows SEM images of the ring resonator with integrated electrodes designed for 4-point resistivity measurements and the Al contact pads for bonding, respectively. Fig.~\ref{F6}(b) shows the I-V characteristics measured on this device at room temperature and at low temperatures. When doped to above the metal-insulator transition, a temperature-independent resistivity is established \cite{Yamanouchi:1967aa}. We investigated devices with peak doping concentrations of both $10^{19}\;\rm cm^{-3}$ and $10^{20}\;\rm cm^{-3}$. Only the $10^{20}\;\rm cm^{-3}$ doped device showed a linear, i.e. metallic, current-voltage trace at room temperature and 4.2~K with a resistivity of $10^{-4}\rm\; \Omega \cdot$cm (assuming a 20 nm thickness) which corresponds to a doping concentration of $4\times10^{19}\;\rm cm^{-3}$. The device doped to a peak concentration of $10^{19}\;\rm cm^{-3}$ was found to be semiconducting at 4.2~K.

\section{Conclusion}
We evaluated the impact of ultra-shallow P doped electrodes on the optical properties of high-Q micro-ring cavity. Modelling results indicate that the loss of the heavily doped Si ($4\times 10^{19} \;\rm cm^{-3}$ within a 20~nm thin layer) waveguide is around $180\rm\; dB/cm$. This value is negligible compared to the coupling loss and transmission loss due to roughness or bending since the implanted area is so small. Experimentally, the doped electrodes in ring-cavity show metallic conductivity at room temperature and 4.2~K and maintain high Q values of the cavity with values of up to several $10^5$. The impact of the fabrication steps related to the electrodes on the Q values is comparable to the variability of the Q values without these additional processing steps. The doped electrodes enable local electrical manipulation of quantum system in silicon that could also be extended to other semiconductor materials such as silicon carbide (SiC), which is important for low-loss processing photons from visible to near infrared wavelengths. 

\section{Acknowledgements}

We thank Dr. Haishu Li and Chunle Xiong for helpful discussions. This work was performed in part at the NSW and ACT Node of the Australian National Fabrication Facility and the University of Sydney nanoscience institute and the School of Physics at the University of Sydney. The facilities as well as the scientific and technical assistance of the Research \& Prototype Foundry Core Research Facility at the University of Sydney are part of the Australian National Fabrication Facility. This research was funded by the Australian Research Council Center of Excellence for Quantum Computation and Communication Technology (No. CE170100012). We acknowledge the AFAiiR node of the NCRIS Heavy Ion Capability for access to ion-implantation facilities.

\bibliography{bib}

\clearpage
\section{Supplementary Material}
\begin{table}[b]
\caption[]{{\bf Resonant mode fitting parameters for fully implanted ring} Spectra appear in Fig. 3 of main text. Split peaks have two sets of values.
} 
\begin{tabular}{l|l|l|l}
Process   & $\lambda$ (nm)                & FWHM (nm)       & Q (10$^4$)       \\\hline\hline
Initial   & 1531.34  /  1531.44   & 0.048  / 0.029  & 3.17 / 5.30     \\
Imp.      & 1529.836              & 0.230           & 0.66            \\
Imp./Ann. & 1528.432              & 0.313           & 0.49            \\\hline
Initial   & 1539.036  /  1539.191 & 0.051 /  0.064  & 3.02 / 2.42     \\
Imp.      & 1537.469  /  1537.604 & 0.185  /  0.190 & 0.83  /  0.81  \\
Imp./Ann. & 1536.109              & 0.286           & 0.54          \\\hline
  Initial   & 1546.887  /  1546.929 & 0.041 /  0.041  & 3.77 / 3.77     \\
Imp.      & 1545.306              & 0.205           & 0.75            \\
Imp./Ann. & 1543.85               & 0.314           & 0.49            \\\hline
 
Initial   & 1554.732  /  1554.806 & 0.048 /  0.048  & 3.24 / 3.21     \\
Imp.      & 1553.15               & 0.230           & 0.67            \\
Imp./Ann. & 1551.64               & 0.267           & 0.58            \\\hline
 Initial   & 1562.657  /  1562.763 & 0.055 / 0.053   & 2.82 / 2.95     \\
Imp.      & 1561.067              & 0.254           & 0.62            \\
Imp./Ann. & 1559.528              & 0.286           & 0.55         \\ \hline
\end{tabular}
\end{table}

\begin{table}[t]
\caption[]{{\bf Resonant mode fitting parameters for locally implanted ring} Spectra appear in Fig. 4 of main text. Split peaks have two sets of values. Before local implantation 3 of the 8 resonant modes observed in the 1530-1570~nm spectral range displayed splitting. After implantation the Q factors generally increased and the split peaks merged together. After annealing the Q factor increased slightly but not up to the initial value. This shows the positive effect of the anneal schedule employed on the resonator quality.
} 
\begin{tabular}{l|l|l|l}
Process   & $\lambda$ (nm)                & FWHM (nm)       & Q (10$^4$)         \\\hline\hline
Initial   & 1530.742              & 0.013           & 11.70           \\
Imp.      & 1531.283  /  1531.318 & 0.015  /  0.015 & 10.10  /  10.47 \\
Imp./Ann. & 1531.373  /  1531.419 & 0.018  /  0.019 & 8.40  /  8.15   \\\hline
Initial   & 1535.586              & 0.013           & 11.43           \\
Imp.      & 1536.151              & 0.016           & 9.47            \\
Imp./Ann. & 1536.254 / 1536.256   & 0.018 / 0.012   & 8.4 / 12.80    \\\hline
Initial   & 1540.437  /  1540.479 & 0.015  /  0.012 & 10.17  /  12.33 \\
Imp.      & 1541.03               & 0.026           & 6.04            \\
Imp./Ann. & 1541.139              & 0.018           & 8.51            \\\hline
Initial   & 1545.329  /  1545.379 & 0.014  /  0.015 & 11.40  /  10.20 \\
Imp.      & 1545.932              & 0.024           & 6.53            \\
Imp./Ann. & 1546.043              & 0.022           & 7.04            \\\hline
Initial   & 1550.266  /  1550.292 & 0.014 /  0.014  & 11.14 /  10.71  \\
Imp.      & 1550.866              & 0.018           & 8.51            \\
Imp./Ann. & 1550.98               & 0.015           & 10.02           \\\hline
Initial   & 1555.228              & 0.013           & 12.07           \\
Imp.      & 1555.822              & 0.019           & 8.28            \\
Imp./Ann. & 1555.936              & 0.025           & 6.18            \\\hline
Initial   & 1560.203              & 0.021           & 7.41            \\
Imp.      & 1560.809              & 0.029           & 5.45            \\
Imp./Ann. & 1560.93               & 0.030           & 5.26            \\\hline
Initial   & 1565.212              & 0.017           & 9.31            \\
Imp.      & 1565.828              & 0.024           & 6.55            \\
Imp./Ann. & 1565.947              & 0.022           & 7.16           \\\hline
\end{tabular}
\end{table}
\end{document}